\documentclass[letterpaper, 10 pt, conference]{ieeeconf}  % Comment this line out if you need a4paper
\IEEEoverridecommandlockouts                              % This command is only needed if 
\usepackage{amsmath} % assumes amsmath package installed
\usepackage{amssymb} % assumes amsmath package installed
\usepackage{float}
\usepackage[pdftex]{graphicx}
\usepackage{tabularx}
\usepackage{verbatim}
\usepackage{color}
\usepackage{xparse}
\usepackage{hyperref}
\usepackage{xmpmulti}
\usepackage{transparent}
\usepackage{fancyhdr}
\usepackage{cite}
\usepackage{rotating}
\usepackage{bbm}
\usepackage{amssymb,amsmath,float}
\usepackage[linesnumbered, algoruled, vlined]{algorithm2e}
\usepackage{booktabs}
\usepackage{caption}
\usepackage{subcaption}
\usepackage{tikz}
 
\usepackage{amsthm}
\usepackage{hyperref}
\usepackage{wrapfig}
\usepackage{setspace}
\usepackage{graphicx}
\usepackage{epsfig}
\usepackage[all]{xy}
\usepackage{verbatim}
\usepackage{float}
\usepackage{booktabs}
\usepackage{multirow}
\usepackage{color,soul}
\usepackage{amssymb,amsmath,graphicx}
\usepackage[compact]{titlesec}
\usepackage{diagbox}
\usepackage{array}
\usepackage{units}
\usepackage[export]{adjustbox}
\DeclareMathOperator{\sech}{sech}

\title{\LARGE \bf
Solving Two-Player General-Sum Game Between Swarms
}

\author{Mukesh Ghimire$^{1}$, Lei Zhang$^{1}$, Wenlong Zhang$^{2}$, Yi Ren$^{1}$ and Zhe Xu$^{1}$ % <-this % stops a space
\thanks{*This work was not supported by any organization}% <-this % stops a space
\thanks{$^{1}$M. Ghimire, L. Zhang, Y. Ren, and Z. Xu are with the School of Engineering, Matter, Transport, and Energy, Arizona State University, Tempe, AZ 85287, USA. Email:
        {\tt\small \{mghimire, lzhan300, yiren, xzhe1\}@asu.edu}}%
\thanks{$^{2}$W. Zhang is with the School of Manufacturing Systems and Networks, Arizona State University, Mesa, AZ 85212, USA. Email:
        {\tt\small wenlong.zhang@asu.edu}}%
}

\begin{document}

\maketitle
\thispagestyle{empty}
\pagestyle{empty}

%%%%%%%%%%%%%%%%%%%%%%%%%%%%%%%%%%%%%%%%%%%%%%%%%%%%%%%%%%%%%%%%%%%%%%%%%%%%%%%%
\begin{abstract}
% \zx{What does data incentive mean?} resolved
Hamilton-Jacobi-Isaacs (HJI) PDEs are the governing equations for the two-player general-sum games. Unlike Reinforcement Learning (RL) methods, which are data-intensive methods for learning value function, learning HJ PDEs provide a guaranteed convergence to the Nash Equilibrium value of the game when it exists. However, a caveat is that solving HJ PDEs becomes intractable when the state dimension increases. To circumvent the curse of dimensionality (CoD), physics-informed machine learning methods with supervision can be used and have been shown to be effective in generating equilibrial policies in two-player general-sum games. In this work, we extend the existing work on agent-level two-player games to a two-player swarm-level game, where two sub-swarms play a general-sum game. We consider the \textit{Kolmogorov forward equation} as the dynamic model for the evolution of the densities of the swarms. Results show that policies generated from the physics-informed neural network (PINN) result in a higher payoff than a Nash Deep Q-Network (Nash DQN) agent and have comparable performance with numerical solvers.
% We compare the policies generated from a NashDQN agent with the policies obtained from solving the HJ PDEs. 

\end{abstract}

%%%%%%%%%%%%%%%%%%%%%%%%%%%%%%%%%%%%%%%%%%%%%%%%%%%%%%%%%%%%%%%%%%%%%%%%%%%%%%%%
\section{INTRODUCTION}

Swarms, or groups of robots have the ability to carry out complex tasks that are difficult for a single agent. Swarms have been deployed to perform diverse tasks, for example, surveillance and reconnaissance in a military capacity, search and rescue, and even entertainment in the form of light shows~\cite{swarm_behavior}. Other applications include task allocations~\cite{task}, complex formation~\cite{formation}, smart farming~\cite{farming}, and decision making~\cite{decision}. As new applications emerge every day, it is imminent that there will be multiple groups of swarms that may be operating in the same region with their own objective. One of the most trivial examples would be a zero-sum game with two swarm groups with competing objectives. Most of the work in swarms focuses on studying swarm behaviors in a more single-agent setting, where interactions are limited within the swarm group. While some work have proposed game theoretic frameworks to solve problems in communication networks~\cite{network_1, network_2}, they do so in a distributed fashion. We are interested in finding a high level control strategy for swarms with arbitrarily large population. 

Modeling of swarms can be broadly divided into \textit{macroscopic} and \textit{microscopic} models. \textit{Macroscopic} models are invariant to the number of agents whereas individual-level \textit{microscopic} models change with the number of agents~\cite{berman2020}, and become intractable as the number of agents gets large~\cite{densitycontrol}. As such \textit{macroscopic} models can be thought of as robust models that can be used to study large-scale swarms. A building block for studying such swarms is the \textit{Kolmogorov forward equation} which describes the evolution of the density of a stochastic process. We define density as the ratio of the population of swarms in a region to its total population. Under certain conditions, the \textit{Kolmogorov forward equation} can be applied to encode the macroscopic behavior of a swarm.

Multi-agent interactions can often be represented as zero-sum or general-sum differential games depending on the objective~\cite{icra_23}. The Nash Equilibrial values of these games are viscosity solutions to the Hamilton-Jacobi-Isaacs (HJI) equations~\cite{crandall1983viscosity}. HJI PDEs are conventionally solved on a mesh and often suffer from the curse of dimensionality as the state dimension increases~\cite{mitchell2003overapproximating}. We apply the differential game theory to the density control of multiple swarm groups and study the interaction that emerges. 

In this work, we formulate a general-sum game between swarm groups which are modeled as a density that evolves in continuous time. We use the power of neural network being the universal function approximator~\cite{nn} to learn the nash equilibrial values governed by the HJI equation. Our contribution to the existing research on learning and control of the swarm systems is the introduction of a game-theoretic approach to model interactions among multiple swarm groups. We assess the effectiveness of our approach by comparing its performance against both a widely employed reinforcement learning method for games and a numerical solver. 

% \zx{Write a sentence or two for the results similar with the last sentence of the abstract.}
% \zx{to the existing research on learning and control of swarm systems}. resolved
% \begin{enumerate}
%     \item \textbf{Introduction of a Game-Theoretic Framework}: Our paper pioneers the use of a game-theoretic approach to model interactions among multiple swarm groups. 

%     % \item \textbf{Divergence\zx{Difference} from Conventional Swarm Studies}: In contrast to conventional swarm studies primarily concerned with achieving predefined target distributions, our research focuses on revealing the emergence of swarm behaviors in diverse environments. We do so without imposing any predetermined target distribution, thereby shedding light on a previously unexplored facet of swarm studies. \zx{I think you better talk about the specific contributions in terms of HJ PDE, NN, etc. instead of these vague sentences which do not convey much information.}
%     % \zx{This is vague. Can you be more specific on what this contribution is?}

% \end{enumerate}

\section{RELATED WORK}
% \zx{Break the related work into different paragraphs. Please write a short title for each paragraph of the related work. End each paragraph with why the proposed method is different than the listed related work.}
\noindent\textbf{Swarm Control.}  Control of swarms is of great interest to the research community. An existing challenge in this line of research is developing models and control mechanisms for large-scale swarms~\cite{berman2020}. \cite{densitycontrol} devised an optimal control strategy for controlling a large-scale swarm to a target distribution. \cite{berman_swarm} used a leader-follower framework for herding a robotic swarm to a desired distribution. Another work that is in the spirit of our work is \cite{pmp_swarm}, which applied Pontryagin's Maximum Principle for control of a large-scale robotic population in an optimal control setting. In addition, \cite{network_1} and \cite{network_2} proposed a game-theoretic approach on a graph to solve the \textit{Coverage game}. Another common line of work includes using a Markov chain that models the evolution of the density distribution. \cite{prob_gtl} used the Markov chain and provided a probabilistic control algorithm for swarms of agents subject to some temporal logic specifications. In this paper, we study the interaction between two large-scale swarm groups in an environment where the swarm groups have their own objective. We formulate a general-sum game between these swarm groups and determine their optimal policy. 

\noindent\textbf{Solving HJI PDEs using Deep Learning.} Recent works have considered using autoregressive methods such as physics-informed machine learning to learn values of zero-sum~\cite{deepreach}, and general-sum~\cite{icra_23} differential games. \cite{icra_23} extended \cite{deepreach} from zero-sum game with continuous value function to a general-sum game with discontinuous values with respect to states and time. Nash Equilibrium values of general-sum differential games satisfy HJI PDEs, which makes the residual of HJI PDEs a good candidate for a loss function used in physics-informed neural networks. By minimizing the PDE residuals alongside the boundary conditions, a neural network, often a deep one, is trained to predict the Nash equilibrial values associated with the game. Since Deep Learning methods are well known for their scalability to high dimensional problems (see \cite{dnn-1}, \cite{dnn-2} for details), we leverage this method to solve the problem of learning the value of general-sum games. 

\noindent\textbf{Multi-Agent Reinforcement Learning (MARL). } MARL, unlike single-agent reinforcement learning, addresses the problem of decision-making involving multiple agents that operate in a common environment. Standard Rollout algorithm is used where the problem is reformulated as a single-agent by using a joint action space~\cite{bertsekasrollout}. Standard Rollout algorithm has been extended to Multi-agent Rollout algorithm in order to reduce the complexity arising from the joint action space~\cite{bertsekasmulti}. While most algorithms are in the spirit of optimal control, some works exist that have formulated multi-agent decision-making problems as a game~\cite{nash},~\cite{nashdqn}. \cite{nash} proposed a modified Q-learning method where Q-values are defined on joint action space and are updated following Nash Equilibrial strategies. \cite{nashdqn} extended this idea from a tabular method to a scalable one using the power of deep neural networks. In contrast to the case studies that are discussed in these two related works, our case studies have significantly larger action sets. 

This paper is organized as follows. In Section~\ref{sec:notations}, we present the notations that appear in the paper frequently. In Section~\ref{sec:prelim}, we discuss the fundamental concepts, assumptions, and challenges that are crucial to the development of the paper. We also motivate the problem formulation in this section. In Section~\ref{sec:methods}, we discuss the algorithms that are central to the contributions highlighted in the paper. In Section~\ref{sec:cases}, we test our algorithm on different case studies. Finally in Section~\ref{sec:concl}, we conclude with some limitations of the current work and the possible future directions.

\section{NOTATIONS}\label{sec:notations}

Borrowing notations from graph theory, we denote a directed graph by the tuple $\mathcal{G} = (\mathcal{V}, \mathcal{E})$ containing a set of $M$ vertices, $\mathcal{V} = \{1, \dots, M\}$, and a set of $N_\mathcal{E}$ edges, $\mathcal{E} \subset \mathcal{V} \times \mathcal{V}$, where $e = (i, j) \in \mathcal{E}$ if there is an edge from vertex $i \in \mathcal{V}$ to vertex $j \in \mathcal{V}$. A source map is defined as $S : \mathcal{E} \rightarrow \mathcal{V}$ and a target map as $T : \mathcal{E} \rightarrow \mathcal{V}$ for which $S(e) = i$ and $T(e) = j$, whenever $e = (i, j) \in \mathcal{E}$. The graph $\mathcal{G}$ is said to be \textit{bidirected} if $(v, w) \in \mathcal{E}$ implies that $(w, v) \in \mathcal{E}$ for all $v, w \in \mathcal{V}$.

We follow \cite{icra_23}'s notation system and the implementation therein for HJI equations. Let $\mathcal{X}_i$ and $\mathcal{U}_i$ denote the state and action space respectively for Player $i$. The time-invariant state dynamics of Player $i$ is denoted by $\dot{\mathbf{x}}_i(t) = f(\mathbf{x}_i(t), \mathbf{u}_i(t))$ for $\mathbf{x}_i(t) \in \mathcal{X}_i$ and $\mathbf{u}_i \in \mathcal{U}_i$. Given a time horizon $T$, the instantaneous and terminal losses of Player $i$ are denoted by $l$, and $c(\mathbf{x}_i, \mathbf{x}_{-i})$ respectively. Note that $l$ is a constant in this formulation, however, in general it is a function of state $(l(\mathbf{x}_i, \mathbf{x}_{-i}))$. For a complete-information general-sum differential game between two players, the value function  for each player is $\nu_i(\cdot, \cdot, \cdot) : \mathcal{X}_i \times \mathcal{X}_{-i} \times [0, T] \rightarrow \mathbb{R}$. We will adopt shorthands $f_i$, $l_i$, $c_i$, and $\nu_i$ respectively to denote player-wise dynamics, losses, and the value. We use $\mathbf{a}_i = (a_i, a_{-i})$ to concatenate ego and the other player's variables. $\nabla_x$ denotes partial derivative with respect to $x$.

\section{PRELIMINARIES} \label{sec:prelim}
\noindent\textbf{Swarm Modeling.} We consider two homogeneous sub-swarms, each containing $N$ agents that occupy the regions or vertices $\mathcal{V}$. The swarms evolve in continuous time over this region, which can be denoted by vertices on the graph. We denote the graph by $\mathcal{G} = (\mathcal{V}, \mathcal{E})$, with vertices $\mathcal{V}$ denoting the regions where swarms reside, and $\mathcal{E}$ representing the edges along which the swarms can transition. 

% \zx{number of agents of sub-swarm $i$?} 

Let, $X_v^i(t)$ be the number of agents of sub-swarm $i \in \{1, 2\}$ in the region $v \in \mathcal{V}$. The fraction (or empirical distribution) of the sub-swarms $i$ at location $v \in \mathcal{V}$ at time $t$ is calculated as $\frac{1}{N} X_v^i(t)$. Let $\Omega = \{\mathbf{y} \in \mathbb{R}^n : \sum_{i=1}^n y_i = 1$\}. As $N\rightarrow \infty$, the empirical distribution converges to a deterministic quantity $\mathbf{x}(t) \in \Omega$, which can be used as a state of the sub-swarm $i$.   We denote the state of the sub-swarm $i$ as a vector $\boldsymbol{x}_i(t)$, whose each entry $x_{i, v}(t)$ denotes the fraction of agents in the region $v \in \mathcal{V}$ at time $t$.   We use the \textit{Kolmogorov forward equation}, also known as the \textit{mean-field model} as in ~\cite{berman_swarm} to evolve the state of the sub-swarm: 

\begin{equation}
    \dot{\boldsymbol{x}}(t) = \sum_{e \in \mathcal{E}} u_e(t)\mathbf{B}_e\boldsymbol{x}(t), \quad \boldsymbol{x}(0) = \boldsymbol{x}^0 \in \Omega
\end{equation}
where $\mathbf{B}_e$ are the control matrices with the following entries

\begin{align*}
    B_e^{ij} = \begin{cases}
        -1 & \text{ if } i = j = S(e) \\
        1 & \text{ if } i = T(e), j = S(e) \\
        0 & \text{ otherwise }
    \end{cases}
\end{align*}

% For the experiments in this paper, we consider $u_e \in \{0, 1\}$.

\noindent\textbf{Hamilton-Jacobi-Isaacs Equations.} The Nash equilibrial values of a two-player general-sum differential game, when exist, solve the following HJI equations $(H)$ and satisfy the following boundary condition $(B)$~\cite{starr1969nonzero}:
\begin{equation}\label{Eq:loss}
    \begin{aligned}
    H(\nu_i, \nabla_{\mathbf{x}_i}\nu_i, \mathbf{x}_i, t) := \nabla_t \nu_i + \nabla_{\mathbf{x}_i}^\top \textbf{f}_i - l_i = 0 \\
    B(\nu_i, \mathbf{x}_i) := \nu_i(\mathbf{x}_i, T) - c_i = 0, \quad \text{ for } i = 1, 2.
    \end{aligned}
\end{equation}

The player's policies for ego agent (and other) are derived by maximizing the equilibrial Hamiltonian $h_i(\mathbf{x}_i, \nabla_{\mathbf{x}_i}\nu_i, t) = \nabla_{\mathbf{x}_i}\nu_i^\top f_i - l_i$; $u_i = \arg \max_{u\in\mathcal{U}_i} \{h_i\}$, and $u_{-i} = \arg \max_{u\in\mathcal{U}_{-i}} \{h_{-i}\}$. 

% In the case where the Hamiltonian is linear with respect to the control inputs, pure strategy Nash equilibrium value may not exist, in that case, we solve for a more conservative minimax value. However, mixed strategy Nash equilibrium always exists for a zero-sum game~\cite{mixed_strategy}, which will be explored in future works. 

\noindent\textbf{Pontryagin's Maximum Principle:} We can use the PMP equation to generate open-loop policies, which is often more tractable than solving HJI equations. These open-loop policies can be used as a basis for evaluating the learned closed-loop policies. For a given initial state $(\boldsymbol{x}_1, \boldsymbol{x}_2) \in \mathcal{X}_1 \times \mathcal{X}_2$, we obtain the open-loop policies by solving the following boundary value problem (BVP) according to PMP:

\begin{equation}\label{Eq: pmp}
    \begin{aligned}
        \dot{\mathbf{x}_i} &= \mathbf{f}_i, \quad \mathbf{x}_i[0] = \Bar{\mathbf{x}}_i, \\
        \dot{\boldsymbol{\lambda}}_i &= -\nabla_{\mathbf{x}_i}\mathbf{h}_i, \quad \boldsymbol{\lambda}_i[T] = -\nabla_{\mathbf{x}_i}c_i, \\
        u_i &= \arg \max_{u \in \mathcal{U}_i} \{h_i\},\quad i = 1, 2
    \end{aligned}
\end{equation}

\noindent$\boldsymbol{\lambda}_i$ are the time-dependent co-states for both players concatenated into one. The costates connect PMP and HJI via $\boldsymbol{\lambda}_i = \nabla_{\mathbf{x}_i}\nu_i$. Note that the solutions to Eq. (\ref{Eq: pmp}) are unique to the initial states.

\noindent\textbf{Learning Values of General-sum Differential Games.} Directly learning values of a zero-sum differential game in a self-supervised fashion was first explored in \cite{deepreach}. In this approach, we directly fit a neural network to satisfy the governing HJI PDE. Let $\hat{\nu}_i(\cdot, \cdot, \cdot) : \mathcal{X}_i \times \mathcal{X}_{-i} \times [0, T] \rightarrow \mathbb{R}$ be a neural network parameterized by $\theta$ that approximates $\nu_i$. With $\mathcal{D} = \left\{\left(\boldsymbol{x}_1^{(k)}, \boldsymbol{x}_2^{(k)}, t^{(k)}\right)\right\}_{k=1}^K$ representing the uniform samples in $\mathcal{X}_i \times \mathcal{X}_{-i} \times [0, T]$, the loss function that guides the learning of the general-sum value is: 

\begin{equation}
    \begin{aligned}
    \min_{\theta} L_1 (\hat{\nu_i}; \theta) := \sum_{k=1}^K \left| H\left(\hat{\nu}_i^{(k)}, \nabla_{\mathbf{x}_i}\hat{\nu}_i^{(k)}, \mathbf{x}_i^{(k)}, t^{(k)}\right) \right| \\
    \quad + \;C_1 \left|B\left(\hat{\nu}_i^{(k)}, \mathbf{x}_i^{(k)}\right)\right|
    \end{aligned}
\end{equation}
% \zx{no indent needed for "where", please fix it everywhere}
\noindent where, $\hat{\nu}_i^{(k)}$ is the output of the neural network, 
$\hat{\nu}_i^{(k)} = \hat{\nu}_i\left(x_i^{(k)}, x_{-i}^{(k)}, t^{(k)}\right)$. $C_1$ balances the HJI PDE loss and the boundary loss. 

% To further guide the training to convergence, we add additional boundary loss that fits the gradient of the final value. The loss function transforms into the following: 

% \begin{equation}
%         \begin{aligned}
%     \min_{\theta} L_1 (\hat{\nu_i}; \theta) := \sum_{k=1}^K \left| H\left(\hat{\nu}_i^{(k)}, \nabla_{\mathbf{x}_i}\hat{\nu}_i^{(k)}, \mathbf{x}_i^{(k)}, t^{(k)}\right) \right| \\
%     \quad + \;C_1 \left|B\left(\hat{\nu}_i^{(k)}, \mathbf{x}_i^{(k)}\right)\right| + C_2 \left|\nabla_{\textbf{x}_i}\hat{v}_i^{(k)} - \nabla_{\textbf{x}_i}c_i\right|
%     \end{aligned}
% \end{equation}

% Here, the additional loss parameter is only computed for the samples at the final time. Similar to $C_1$, $C_2$ balances the losses. 

Note that at each training iteration, we compute the control policies for each player by maximizing their equilibrial Hamiltonian. We refer the readers to \cite{icra_23} for more details on challenges and different methods of learning values using this approach. For the purpose of this paper, we only consider the self-supervised learning method to learn the value function.

\section{METHODOLOGY}\label{sec:methods}
In this section, we present the algorithms for generating open-loop trajectories via solving BVP and training the value network using self-supervised learning method. 

\subsection{BVP Solver}
In this subsection, we discuss about generating open-loop trajectories. We use scipy's \texttt{solve\_BVP} function to solve Eq. (\ref{Eq: pmp}). To successfully solve the BVP, we first compute the analytical expressions for the quantities in Eq. (\ref{Eq: pmp}). These are (1) the augmented dynamics which contains the state dynamics $(\dot{\mathbf{x}})$ and the co-state dynamics $(\dot{\boldsymbol{\lambda}})$, and (2) the boundary conditions. Note we also need a sub-routine to compute the control policies for each of the agents. Furthermore, convergence requires good guesses of the solution along the trajectory. To do so, given an initial state, we solve the state and co-state trajectories by solving their respective ODEs obtained from PMP. An alternative to this, albeit slower, is implementing time-marching~\cite{nakamura2021adaptive}. 
% \zx{Is this sentence finished?}

\subsection{Multi-Agent RL using Nash Q-Learning}
We also formulate the 2-regions case as a Reinforcement Learning problem and train the sub-swarms using Nash Q-Learning. Instead of using a tabular method, we use a neural network, especially a Double Deep Q-Network that approximates the Q values. We implement the vanilla Nash Q-Learning following~\cite{nash}. The Nash Q-function for a learning agent $i$ is defined on the state of the system, and its action along with the actions of the other players, whereas in the Q-learning algorithm, the Q-function is only a function of the agent. Furthermore, in Q-learning, agents update their Q-values by accounting for the future values as a result of maximizing their Q-values, whereas in Nash Q-learning, agents update their Q-values by following a Nash equilibrium strategy. More formally, the update is as follows:

\begin{equation}
    \begin{aligned}
        Q^i_{t+1}(s, a^1, \dots, a^n) &= (1-\eta_t)Q^i_t(s, a^1, \dots, a^n) + \\
        & \eta [r^i_t + \beta NashQ^i_t(s')] 
    \end{aligned}
\end{equation}
\noindent where, $\eta$ is the learning rate, and $\beta$ is the discount rate. $Nash$ is the operator of choice that computes the Nash equilibrium of the stage game at state $s'$. In this work, we use \textit{support enumeration} as a nash operator. 

Unlike the vanilla Nash Q-learning, we use two neural networks that approximate the Q-function of the sub-swarms in the 2-regions case. Each sub-swarm has an action space of 4 -- (0, 0), (0, 1), (1, 0), (1, 1). Note that each agent must also know the Q-values of the other agent to compute the Nash Equilibrium. However, this information is not available, so each agent makes a conjecture about the other agent's Q-values. We achieve this by defining the Q-function on the joint action space of both sub-swarms. The result is a $4 \times 4$ matrix of Q-values for each sub-swarm. At each learning stage, each sub-swarm solves a bi-matrix game as follows:

% \usepackage{multirow}
\begin{table}[!ht]
\caption{Bi-matrix stage game for Nash Q-Learning.}
\label{table:bimatrix}
\begin{tabular}{cccccc}
                                                                                               &                             &                                       & \multicolumn{2}{c}{Agent $-i$}                                            &                                       \\ \cline{2-6} 
\multicolumn{1}{c|}{}                                                                          & \multicolumn{1}{c|}{}       & \multicolumn{1}{c|}{(0, 0)}           & \multicolumn{1}{c|}{(0, 1)}           & \multicolumn{1}{c|}{(1, 0)}           & \multicolumn{1}{c|}{(1, 1)}           \\ \cline{2-6} 
\multicolumn{1}{c|}{\multirow{2}{*}{\begin{tabular}[c]{@{}c@{}}Agent\\ $i$\end{tabular}}} & \multicolumn{1}{c|}{(0, 0)} & \multicolumn{1}{c|}{$a_1, b_1$}       & \multicolumn{1}{c|}{$a_2, b_2$}       & \multicolumn{1}{c|}{$a_3, b_3$}       & \multicolumn{1}{c|}{$a_4, b_4$}       \\ \cline{2-6} 
\multicolumn{1}{c|}{}                                                                          & \multicolumn{1}{c|}{(0, 1)} & \multicolumn{1}{c|}{$a_5, b_5$}       & \multicolumn{1}{c|}{$a_6, b_6$}       & \multicolumn{1}{c|}{$a_7,  b_7$}      & \multicolumn{1}{c|}{$a_8, b_8$}       \\ \cline{2-6} 
\multicolumn{1}{c|}{}                                                                          & \multicolumn{1}{c|}{(1, 0)} & \multicolumn{1}{c|}{$a_9, b_9$}       & \multicolumn{1}{c|}{$a_{10}, b_{10}$} & \multicolumn{1}{c|}{$a_{11}, b_{11}$} & \multicolumn{1}{c|}{$a_{12}, b_{12}$} \\ \cline{2-6} 
\multicolumn{1}{c|}{}                                                                          & \multicolumn{1}{c|}{(1, 1)} & \multicolumn{1}{c|}{$a_{13}, b_{13}$} & \multicolumn{1}{c|}{$a_{14}, b_{14}$} & \multicolumn{1}{c|}{$a_{15}, b_{15}$} & \multicolumn{1}{c|}{$a_{16}, b_{16}$} \\ \cline{2-6} 
\end{tabular}
\end{table}
% \zx{Lemke–Howson, also add citations here} resolved
\noindent where, $a$ is the q-value for agent $i$, and $b$ is the q-value for agent $-i$ as conjectured by agent $i$. Methods such as \textit{Lemke-Howson}~\cite{lemke1964equilibrium} or \textit{support enumeration} can be used to compute the Nash Equilibria for the bi-matrix game. Note that  we use \textit{NashPy}'s implementation of \textit{support enumeration}~\cite{nashpyproject}. Note that table~\ref{table:bimatrix} gets significantly larger when we consider higher dimensional case studies. Solving the bi-matrix game becomes a challenge for these cases. As a result we only compare the performance of Nash DQN with PINN for the case study with 2 regions. We describe the implementation details for Nash DQN briefly in Algorithm~\ref{algo:nashddqn}.

\begin{algorithm}[!ht]
\SetAlgoLined
\caption{Nash DQN}
\label{algo:nashddqn}
\SetKwInOut{Input}{input}
\SetKwInOut{Output}{output}
\SetKwInOut{Initialize}{initialize}
\Input{\# Episodes $B > 0$, Minibatch size $m > 0$,  Episode length $N$, Exploration rate $\varepsilon$, Discount rate $\beta$, soft-update rate $\tau$}
\Initialize{Replay Buffer $\mathcal{D}$, Q-networks $(Q_{\theta_1}, Q_{\theta_2})$, and respective target networks $(Q_{\theta_{1'}}, Q_{\theta_{2'}})$}
\For{Episode  b $\leftarrow$ 1 to B}{
    Reset Simulation, sample initial state $s$ randomly.
    
    \For{Episode steps t $\leftarrow$ 1 to N}{
    Select actions $u_1, u_2$ via $NashQ_{\theta_i}(s)$ with $\varepsilon$ random exploration\;
    Store $\mathcal{D} \leftarrow (s, u_1, u_2, r_1, r_2, s')$\;
    Sample transitions $\mathcal{Y} = \{y\}_{i=1}^m$ from $\mathcal{D}$\;
    Set $Q_i \leftarrow Q_{\theta_i}(s, u_1, u_2), i=\{1, 2\}$\;
    % Compute $u_{i'}$ via $Nash Q_i$ (support enumeration), $i = \{1, 2\}$\;
    Perform gradient descent steps on $(Q_i(s, u_1, u_2) - [r_i + \beta Nash Q_{\theta_{i'}}(s')])^2$, $i = \{1, 2\}$\;
    Update target network parameters: $$\theta_{i'} \leftarrow \tau \times \theta_i + (1-\tau) \times \theta_{i'}$$
    }
}
\end{algorithm}

\subsection{Self-supervised learning}

In this subsection, we discuss the method of self-supervised learning algorithm using PINN to learn the value function. We implement curriculum learning by first learning the value at the final time and gradually increasing the time horizon starting from the end time. We provide a simplified algorithm for training the value network. In our experiment, we use ADAM to optimize the neural network parameter $(\theta)$ with a decaying step-size (learning rate) $\gamma \in [2e-5, 1e-6]$. We use a neural network with 3 hidden layers containing 64 neurons each with $\texttt{tanh}$ activation function, with the final layer being linear. For the higher dimensional case study we increase the hidden layer to 5 and the neurons to 128 at each layer. We use NVIDIA A100 with 2 GPUs for all the training in this paper. The dimension of input $\mathcal{X}$ to the neural network depends on the number of regions in the graph. We further reduced the dimension of the system using the fact that the densities of each sub-swarm sum to 1. Hence, for the case with 2 regions, the value function is a 3-dimensional function instead of 5, including time. 
% \zx{Be specific. densities of each sub-swarm in the sub-regions?}
\begin{algorithm}[!ht]
\SetAlgoLined
\caption{Self-supervised Learning}
\SetKwInOut{Input}{input}
\SetKwInOut{Output}{output}
\Input{$T$, $num\_epoch$, $pretrain\_iters$, $k$, $n$}
\Output{$V_\theta$}
initialize neural network $V_\theta$\;
sample $k$ samples of $\mathcal{X} \in \Omega$\;
pre-train $V_\theta$ at the boundary $(t=0)$ for $pretrain\_iters$ iterations\;
set $itr = 0$\;
\While{$itr \le num\_epoch$}{
    sample $k$ samples of $\mathcal{X} \in \Omega$\;
    sample $k$ samples of $t \sim \mathcal{U} (0,  T (itr)/{num\_epoch})$\;
    append $n$ samples at the boundary $(t = 0)$ to $\mathcal{X}$ and to $t$\;
    compute total loss ($L_1$) using PDE loss $(H)$, and boundary loss $(B)$\;
    update $\theta \leftarrow \theta + \gamma \nabla_{\theta}L_1$
}
\end{algorithm}

% uniform_(0, (self.t_max - self.t_min) *
% (self.counter / self.full_count))

\section{CASE STUDIES}\label{sec:cases}
\noindent\textbf{Objective.} Now, let us define the general-sum game that we study in this paper along with the case studies. Formally, the payoff and the value of the game for a player $i$ is defined as: 

% \zx{Better write out the full function for the last line, $J_i(\boldsymbol{x}_i(0), \boldsymbol{x}_{-i}(0), u_i, u_{-i})$ instead of $J_i$} 
\begin{align}
    J_i(\boldsymbol{x}_i(0), \boldsymbol{x}_{-i}(0), u_i, u_{-i}) &=  g(\boldsymbol{x}_i(T), \boldsymbol{x}_{-i}(T)) \nonumber \\
    \dot{\boldsymbol{x}}_i &= f(\boldsymbol{x}_i, u_i)\\
    V_i^* &= \max_{u_i} J_i(\boldsymbol{x}_i(0), \boldsymbol{x}_{-i}(0), u_i, u_{-i}) \nonumber
\end{align}

Note that we do not have an instantaneous loss in the definition of the value, $(l_i = 0, \forall i)$. The dynamics equation, $f$ is different for the case studies, which depends on the underlying graph. 

\begin{figure}[!h]
    \centering
    \includegraphics[width=\linewidth]{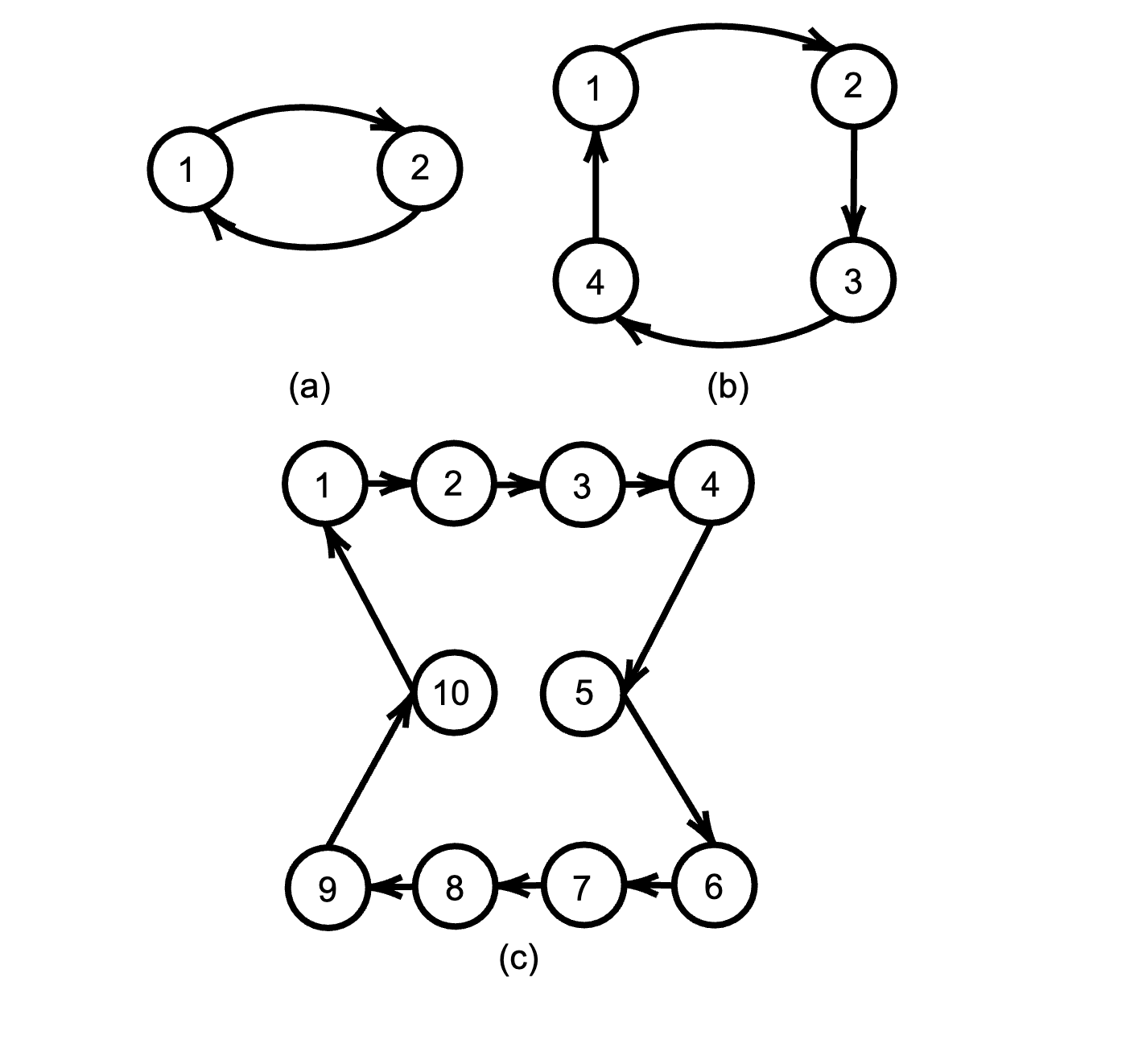}
    \caption{(a) Bidirected graph with 2 regions or vertices. (b) Directed graph with 4 regions. (c) Directed graph with 10 regions. The maximum transition rates (control bounds) for all the edges is a constant 1, and the minimum rate is 0.}
    \label{fig:cases}
\end{figure}

We test our method in three different graphs representing both low and high dimensional systems. We consider two sub-swarms with the same control capabilities interacting in an environment containing $M$ regions $(|\mathcal{V}| = M)$. We first consider $M=2$ (as a toy case) and $M=4$ regions with the former represented by a bidirected graph and the latter represented by a directed graph as shown in Fig.~\ref{fig:cases}. We then consider a high dimensional case (21 D), with $M=10$ regions, to show that the proposed method is easily scalable. 

The state of each sub-swarm is denoted by $\mathbf{x}_i \in \mathbb{R}^M$, with $M$ being the number of regions. The goal of each sub-swarm is to achieve higher proportion in any one of the regions (vertices in the graph). The terminal payoff to sub-swarm $i$ is $g(\mathbf{x}_i, \mathbf{x}_{-i})$ is a Boltzmann operator $(S_{\alpha})$ defined on the element-wise difference between the states of the two sub-swarms. We use the Boltzmann operator instead of the $\max$ operator to make the function differentiable and continuous. The following represents the terminal payoff to the sub-swarm group $i=1$. To compute the terminal payoff for sub-swarm group $i=2$, simply reverse the order of difference. 
\begin{equation}
    \begin{aligned}
        g(\mathbf{x}_1, \mathbf{x}_2) = S_{\alpha}(\mathbf{x}_1 - \mathbf{x}_2) = \frac{\sum_{j=1}^M \left(\mathbf{x}_1 - \mathbf{x}_2\right)_j e^{\alpha (\mathbf{x}_1 - \mathbf{x}_2)_j}}{\sum_{j=1}^M e^{\alpha (\mathbf{x}_1 - \mathbf{x}_2)_j}} 
    \end{aligned}
\end{equation}
where $\alpha$ is the temperature parameter and as $\alpha \rightarrow \infty$, it behaves like the $\max$ operator. We set $\alpha = 1$ for the experiments in this paper (unless mentioned explicitly). The gradient of $S_\alpha$ resembles that of a \texttt{softmax}:

\begin{equation}
    \begin{aligned}
        \nabla_{{x}_i} S_\alpha (x_1, \dots, x_M) &= \frac{e^{\alpha}x_i}{\sum_{j=1}^M e^{\alpha x_j}}\Big[1 + \\
        &\quad \quad \alpha(x_i - S_\alpha(x_1, \dots, x_M))\Big]
    \end{aligned}
\end{equation}

An important thing to note is that the states in our formulation are not independent as they must sum to 1. We highlight how this brings about some crucial changes to the formulation using the 2-regions case study. After reducing the state dimension from two-dimensional to one, denote the state of sub-swarm 1 with $x$ and that of sub-swarm 2 with $y$, which represents their density in region 1. Then, the final payoff becomes:
\begin{align*}
    c_{i=1} = S(x-y) &= \frac{(x-y) e^{(x-y)} + (y-x) e^{(y-x)}}{e^{(x-y)} + e^{(y-x)}}\\
                     &= (x-y)\frac{e^{(x-y)} - e^{-(x-y)}}{e^{(x-y)} + e^{-(x-y)}} \\
                     &= (x-y)\tanh{(x-y)}
\end{align*}
And, its gradient with respect to the state variables are:
\begin{align*}
    \frac{\partial c_{i=1}}{\partial x} &= \tanh{(x-y)} + (x-y)\sech^2(x-y)\\
    \frac{\partial c_{i=1}}{\partial y} &= - \frac{\partial c_{i=1}}{\partial x}
\end{align*}
% These calculations are necessary in order to obtain the BVP solutions and are not required for neural network training. 

% Now let us define the general-sum game that we study in this paper. Formally, the payoff and the value of the game for a player $i$ is defined as:
% \begin{align*}
%     J_i(\boldsymbol{x}_i(0), \boldsymbol{x}_{-i}(0), u_i, u_{-i}) &=  g(\boldsymbol{x}_i(T), \boldsymbol{x}_{-i}(T)) \\
%     \dot{\boldsymbol{x}}_i &= f(\boldsymbol{x}_i, u_i)\\
%     V_i^* &= \max_{u_i} J_i  
% \end{align*}

% Note that we do not have an instantaneous loss in the definition of the value, $(l_i = 0, \forall i)$.
\subsection{Sub-swarms in 2 regions}

\begin{figure}[!ht]
    \centering
    \includegraphics[scale=0.35]{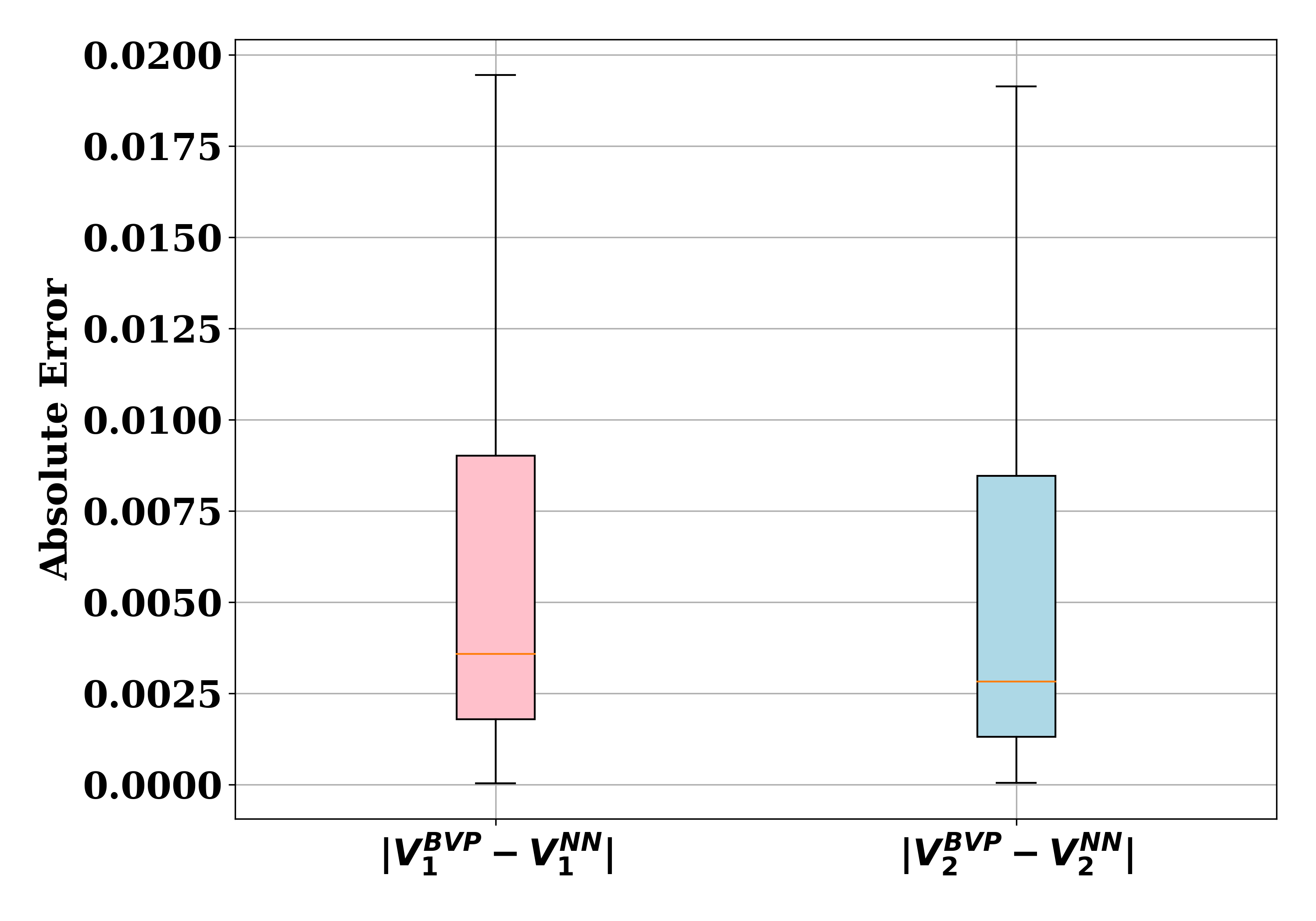}
    % \vspace{-0.139in}
    \caption{Absolute Error of Values at final time starting from same initial states between BVP solution and neural network. $(V_1^{BVP}$, $V_2^{BVP})$ represent values of sub-swarms 1 and 2 obtained from BVP solver, and $(V_1^{NN}, V_2^{NN})$ are obtained from value network.}
    \label{fig:BVP-nn}
\end{figure}

We first solve a simple case with only 2 regions and identical homogeneous sub-swarms. We compare the open-loop policy obtained by solving the BVP using Eq. (\ref{Eq: pmp}) with the policy obtained from the neural network. First, we show that the closed-loop policy from the neural network results in approximately the same final payoff for the swarm groups. We take 1000 initial states and find the absolute error between the values at the final time both using the BVP solver and the neural network. Fig.~\ref{fig:BVP-nn} shows that the resulting payoffs are close to each other. 

\noindent\textbf{Training.} For 2 dimensional state space, we sample 65k data and pretrain at the boundary for 10k iterations. We then implement curriculum learning by slowly increasing the time horizon and train for another 110k iterations. We also ensure that there are at least 10k data points at the boundary after the pretrain stage. 

% First, we show that the neural network approximation of the value is close to the analytical solution obtained from solving the BVPs. 

\begin{figure}[!htb]
    \centering
    \includegraphics[width=\linewidth]{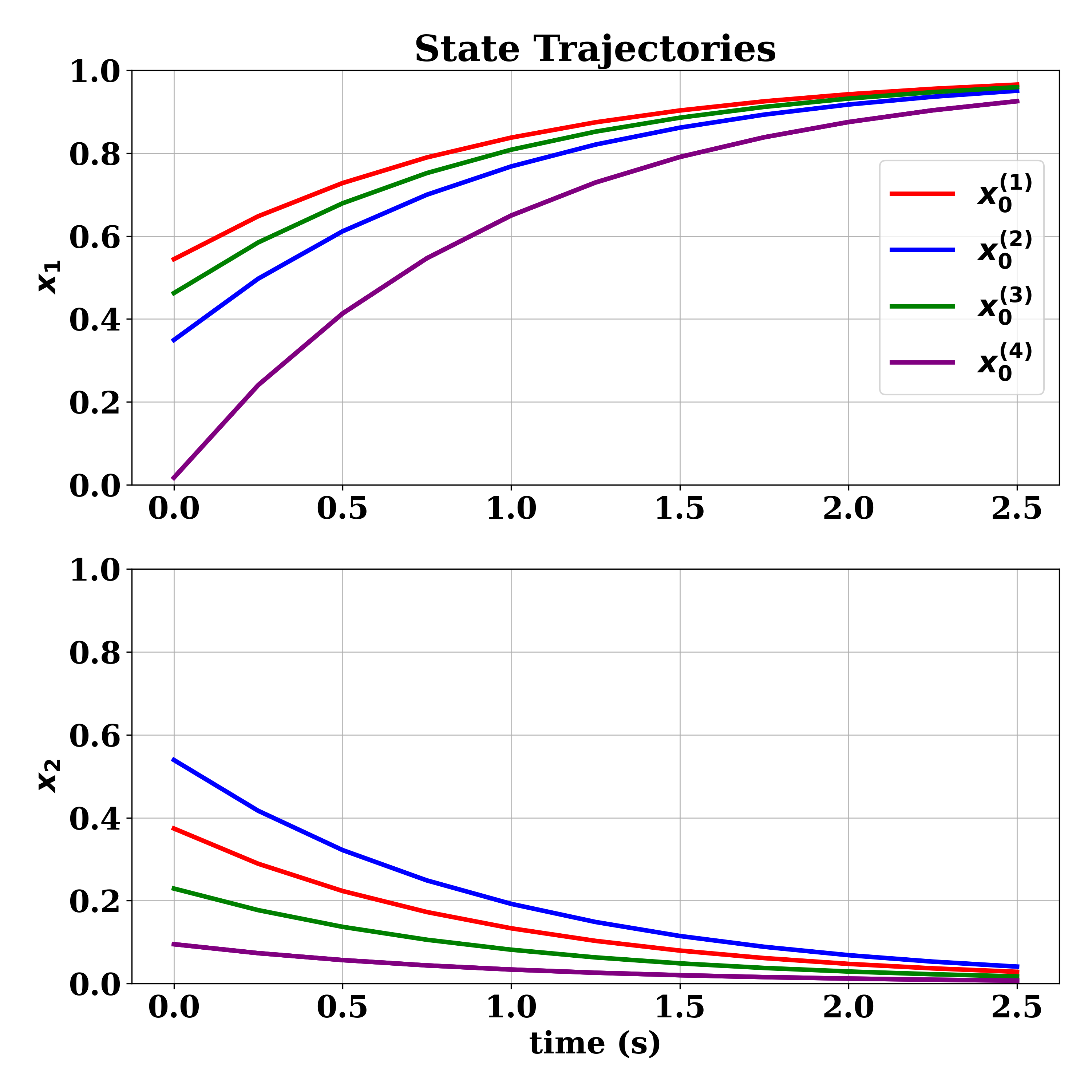}
    \caption{Interaction in two regions. Density evolution of swarm groups starting from four different initial conditions, $\textbf{x}_0^{(1)}$, $\textbf{x}_0^{(2)}$, $\textbf{x}_0^{(3)}$, and $\textbf{x}_0^{(4)}$, which represent the densities of the swarm groups in region 1. The top figure corresponds to sub-swarm 1 and the bottom corresponds to sub-swarm 2. Each sub-swarm aggregates in one region each, which maximizes their payoffs.}
    \label{fig:swarm_2d}
\end{figure}

\noindent\textbf{Results.} Using the learned value network, we generate the closed-loop trajectories for a given initial condition representing the density of swarm groups in region 1. The results show that each swarm group will occupy one separate region at the end of the game. Fig.~\ref{fig:swarm_2d} shows the trajectories of the swarm groups (top corresponds to Swarm Group 1), over the time horizon of 2.5 seconds starting from 4 different initial conditions. This solution maximizes payoffs of both swarm groups and therefore is one of the Nash Equilibrium solutions. 

\begin{figure}
    \centering
    \includegraphics[width=\linewidth]{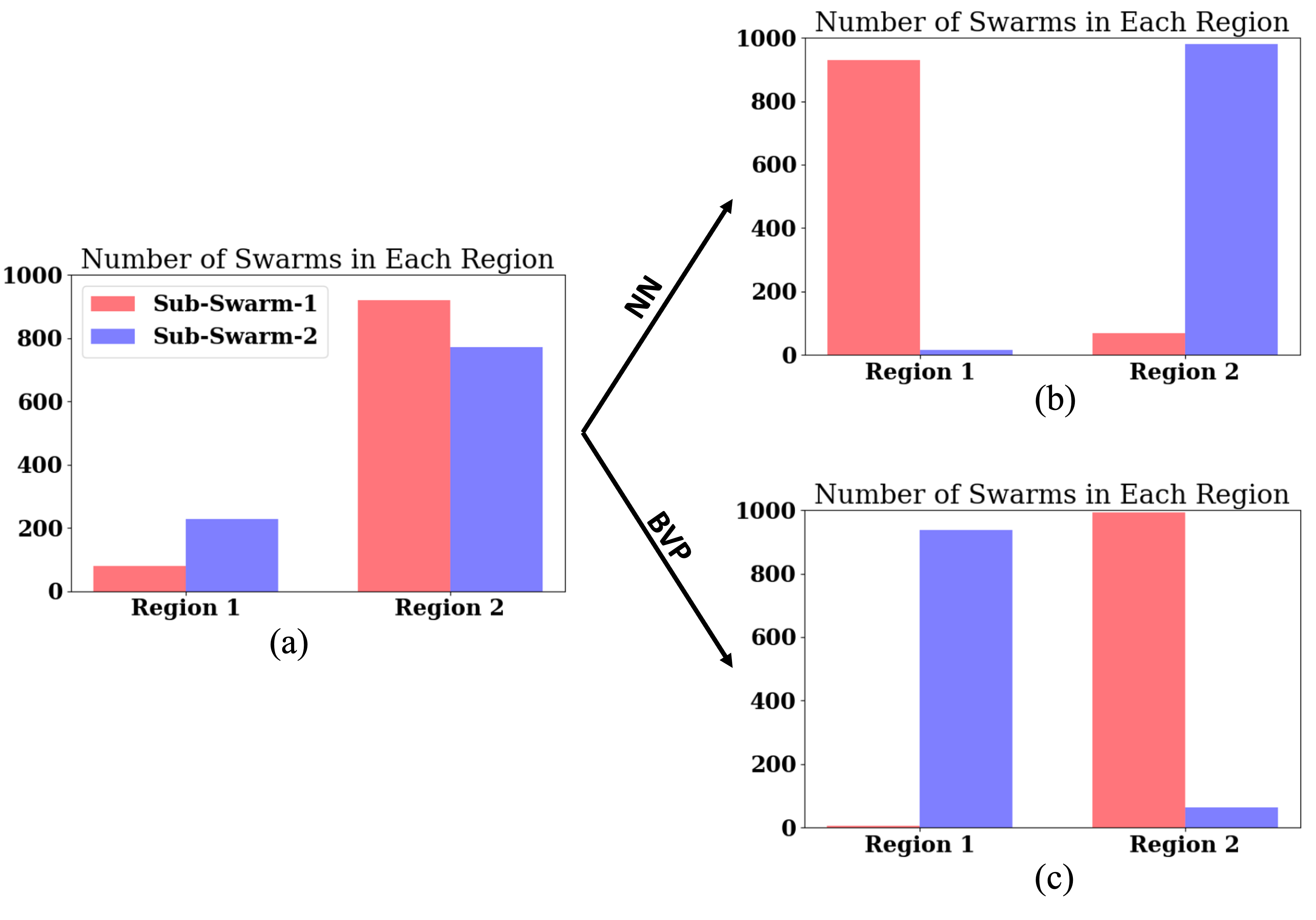}
    \caption{A case demonstrating possible equilibria for the 2 regions case. (a) is the initial condition of the sub-swarms. (b) and (c) are the final distributions of the sub-swarms in the environment obtained from the value network and the BVP solver respectively. Both distributions are equilibria for the sub-swarms as they result in approximately equal payoffs. Multiple equilibria exist due to the symmetric nature of the game.}
    \label{fig:multi_eq}
\end{figure}

We also observe that for some initial states value network and the BVP solver result in different equilibria. Since the problem is symmetric, it is possible that the swarms can occupy any one of two regions to guarantee a maximum payoff as long as they pick separate regions. We show one example of such case in Fig.~\ref{fig:multi_eq}. To further test that these solutions are indeed a Nash Equilibrium, we generate 100 different trajectories using the learned value network as a closed-loop controller. These trajectories are then fed into the BVP solver as an initial guess. The resulting solution from the BVP solver converged to the solution obtained from the neural network for 97 out of the 100 total trajectories. 

\subsection{Sub-swarms in 4 regions}
Next, we consider an environment with 4 regions as shown in Fig.~\ref{fig:cases}(c). Number of swarm groups is 2 with same transition rates across all edges. 

\noindent\textbf{Training.} For 4 regions case, we increase the training iterations to 200k with 20k pretrain steps. In addition to curriculum learning, we also increase the data points gradually starting from 30k in the beginning and adding 10k data points after every 10k iterations after the pretrain stage.

\begin{figure}[!htb]
    \centering
    \includegraphics[width=\linewidth]{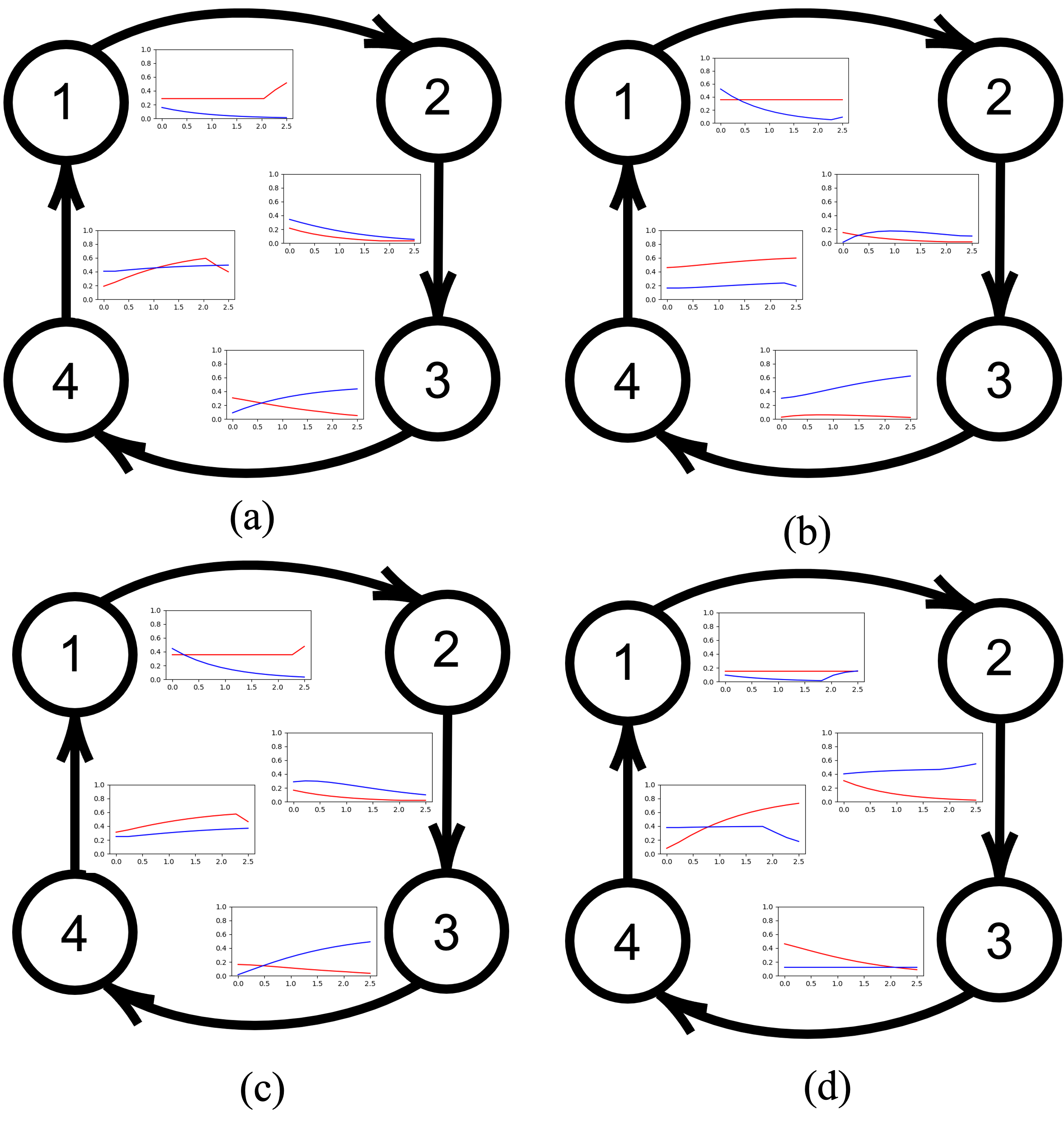}
    \caption{Interaction in four regions. Density evolution of swarm groups from four different initial conditions (a - d). Red plot corresponds to the density of sub-swarm 1 at that particular region, and the blue corresponds to that of sub-swarm 2. Sub-Swarms tend to find at least one region where they maintain dominance.}
    \label{fig:4d-graph}
\end{figure}

\begin{figure}[!ht]
    \centering
    \includegraphics[width=\linewidth]{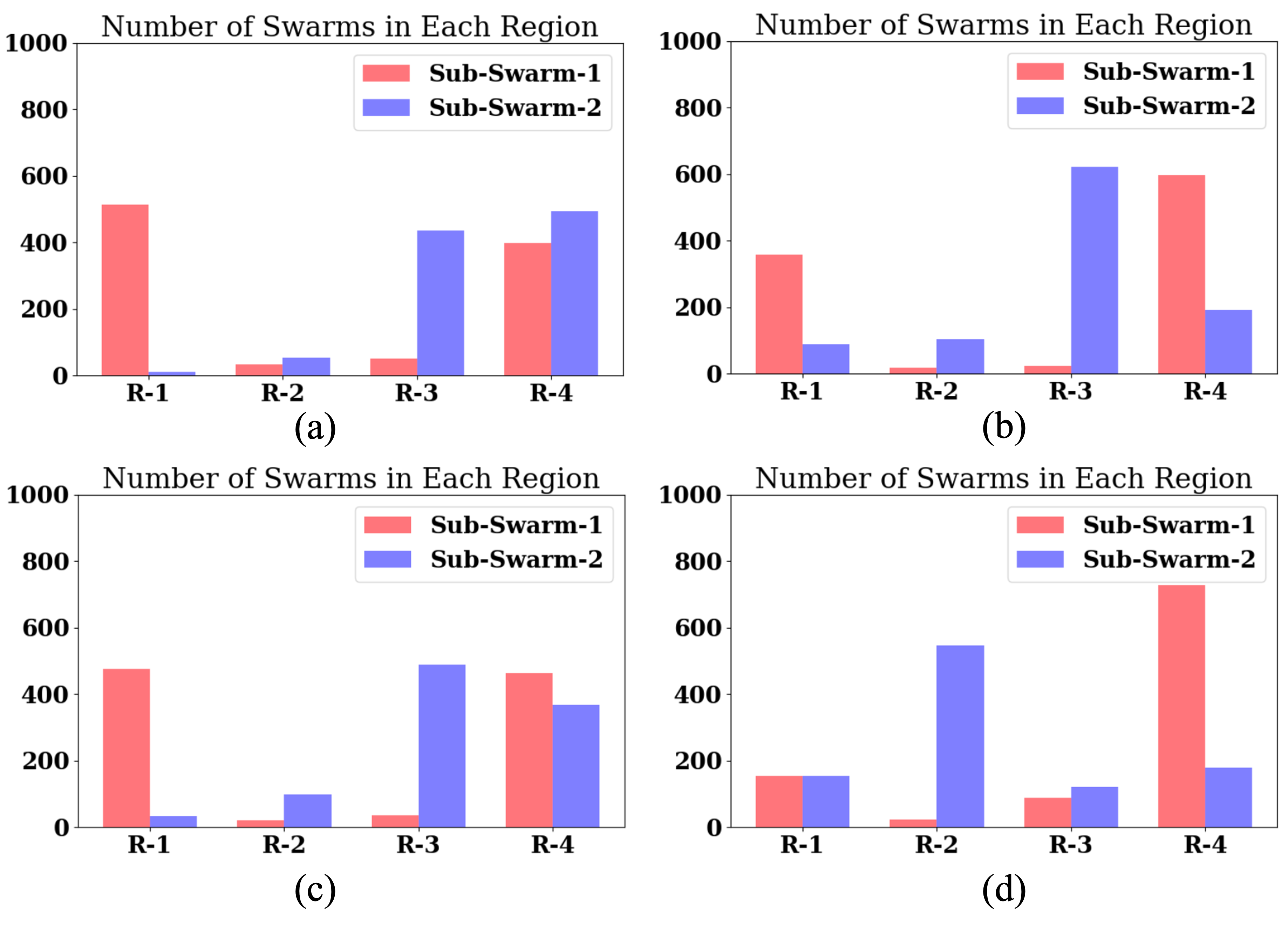}
    \caption{(a-d) Represents the density distribution for two sub-swarms at the final time for four different initial conditions. The initial conditions are the same as in Fig.~\ref{fig:4d-graph}.}
    \label{fig:4d_density}
\end{figure}

\noindent\textbf{Results.}
As in the previous case study, the sub-swarms eventually find one region each where they dominate. Fig.~\ref{fig:4d-graph} shows the evolution of densities at the four regions for four different initial conditions. The densities at the final time for these four initial conditions are shown in Fig.~\ref{fig:4d_density}. All cases lead to the scenario where sub-swarms maintain dominance in one region. 

We use the initial states of the 150 closed-loop trajectories generated using the value network as a controller and solve the game using the BVP solver. We then compare the open-loop trajectories of the converged cases with the initial states from the closed-loop trajectories and report the final payoff differences. Note that some initial states when solved using the BVP solver may not converge to a solution. While, for the 2 regions case, all 150 initial states achieved convergence, for the 4 regions case, 139 out of the total 150 initial states reached convergence.

\begin{figure}
    \centering
    \includegraphics[width=\linewidth]{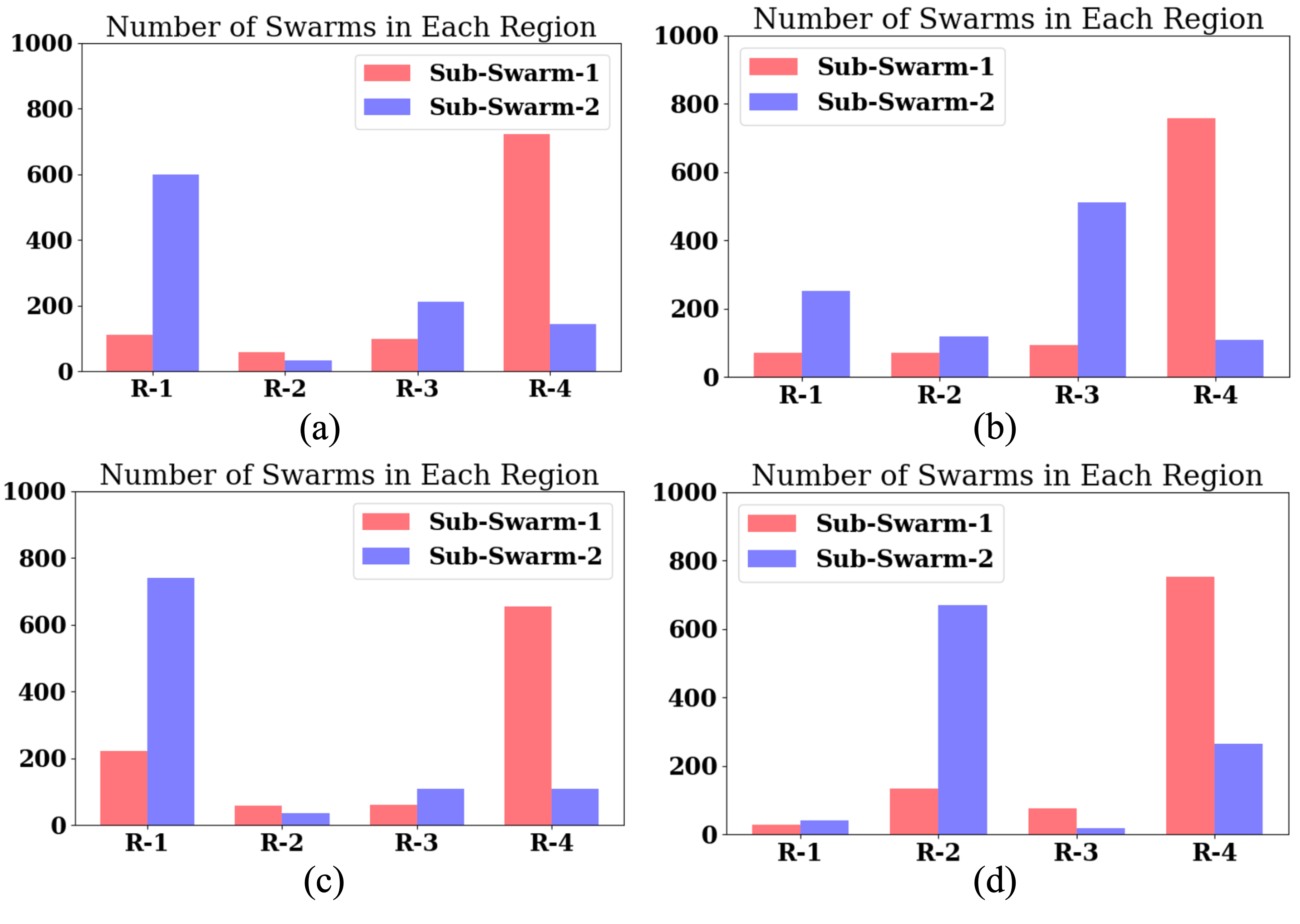}
    \caption{Final distributions of the sub-swarms when the temperature parameter $\alpha=20$. For the same initial conditions as in the previous cases, the sub-swarms attempt to aggregate in one region, which is desirable.}
    \label{fig:alpha_20}
\end{figure}

We observe that for some initial states, the final distributions are not optimal (see Fig.~\ref{fig:4d_density}(a) and (c)). Sub-swarm-2 in Fig.~\ref{fig:4d_density}(a) could get a higher payoff by aggregating in region 3 instead of two regions (3 and 4). This behavior can be attributed to the temperature parameter in the payoff function. Since hardening the temperature parameter led to difficulty in the convergence of BVP, we tested the effect of the temperature parameter using the value network. We trained a value network with $\alpha=20$ in the final payoff function. The results show that the sub-swarms proceed to aggregate in one region resulting in the desired behavior. Fig~\ref{fig:alpha_20} shows the final distributions of the sub-swarms for the same four initial conditions mentioned before.

We compare the final payoffs across the learning algorithms for both 2 and 4 regions case studies. The results are reported in table~\ref{table:payoff}. A negative number means that the BVP solution resulted in a higher final payoff than the respective learning method. Final payoffs using PINN were close to the that obtained from the BVP solution whereas Nash DQN resulted in a significantly lower payoff on average. 

% Next, we increase the state dimension from 4 to 8 and study the behavior of the swarm using the approximation of the value function obtained from the neural network. 
% \usepackage{booktabs}
% \usepackage{multirow}
% \usepackage{booktabs}
% \usepackage{multirow}
% \usepackage{booktabs}
% \usepackage{multirow}
% \usepackage{booktabs}
% \usepackage{multirow}
\begin{table}[!ht]
\scriptsize
\caption{Comparison of final payoffs from solving the interaction using physics-informed neural network and Nash-DQN with BVP solver. $\overline{V_1}$ denotes mean final payoff to sub-swarm 1.} 
\label{table:payoff}
\centering
\begin{tabular}{@{}cccccc@{}}
\toprule
\multicolumn{1}{l}{Environment}                                                & \multicolumn{1}{l}{$\overline{V_1}$} & \multicolumn{1}{l}{$\overline{V_2}$} & \multicolumn{1}{l}{Algorithm}                      & \multicolumn{1}{l}{$\overline{\hat V_1 - V_1}$} & \multicolumn{1}{l}{$\overline{\hat V_2 - V_2 }$} \\ \midrule
\multirow{2}{*}{\begin{tabular}[c]{@{}c@{}}Case-1 \\ (2 regions)\end{tabular}} & \multirow{2}{*}{0.697}   & \multirow{2}{*}{0.697}   & PINN                                               & \textbf{-0.011}                          & \textbf{-0.012}                           \\\cmidrule(l){4-6} 
                                                                               &                          &                          & \begin{tabular}[c]{@{}c@{}}Nash\\ DQN\end{tabular} & -0.162                           & -0.281                           \\ \midrule
\begin{tabular}[c]{@{}c@{}}Case-2 \\ (4 regions)\end{tabular}                  & 0.150                   & 0.147                   & PINN                                               & \textbf{0.0001}                     & \textbf{-0.002}                           \\ \bottomrule
\end{tabular}
\end{table}

\subsection{High Dimensional Case - 10 regions}
Finally, to end the paper, we present a higher dimensional case study as shown in Fig.~\ref{fig:cases}(c). There are 10 regions connected by a directed graph with 10 edges. Each sub-swarm group has to pick a 10-dimensional action at each time step. The action space of each sub-swarm is $2^{10}$, which makes the problem intractable to solve in the case of Nash Q-learning and BVP. Solving BVP becomes intractable as it requires computing guesses for each of the actions. 

\noindent\textbf{Training.} We slightly increase the neural network size to 5 hidden layers containing 128 neurons each. We keep other hyper-parameters the same as those in the 4 regions case. 

\begin{figure}[!htb]
    \centering
    \includegraphics[width=\linewidth]{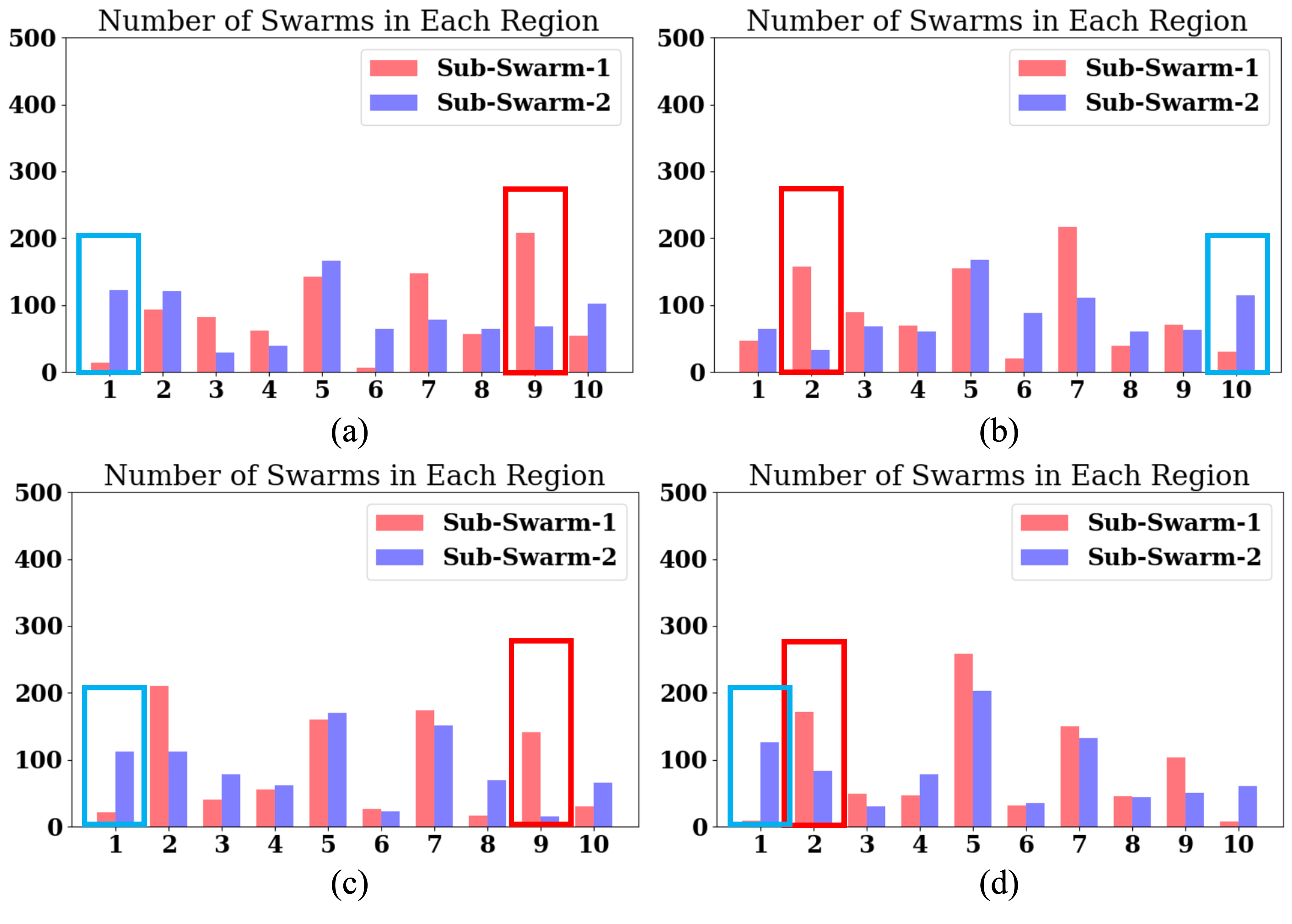}
    \caption{(a-d) Shows the density distribution of two sub-swarms at the final time in the environment with 10 regions. Similar to the 4 regions case, the sub-swarms are able to find a region where they dominate. The regions where the sub-swarms dominate are bounded with a box.}
    \label{fig:highdim}
\end{figure}

\noindent\textbf{Results.}
The final distributions of the sub-swarms across the 10 regions for 4 different initial conditions are shown in Fig.~\ref{fig:highdim}. From the plot, we see that there exists at least one region for both sub-swarms where their proportions are higher compared to the other sub-swarm. We do not observe a pronounced difference in densities of the sub-swarms as in the case of 4 and 2 regions. A possible explanation for not achieving this behavior could be an insufficient time horizon. It is possible that the time horizon of 2.5 seconds is not enough for sub-swarms to be able to aggregate in one region.

%    \begin{figure}[thpb]
%       \centering
%       \framebox{\parbox{3in}{We suggest that you use a text box to insert a graphic (which is ideally a 300 dpi TIFF or EPS file, with all fonts embedded) because, in an document, this method is somewhat more stable than directly inserting a picture.
% }}
%       %\includegraphics[scale=1.0]{figurefile}
%       \caption{Inductance of oscillation winding on amorphous
%        magnetic core versus DC bias magnetic field}
%       \label{figurelabel}
%    \end{figure}

% Figure Labels: Use 8 point Times New Roman for Figure labels. Use words rather than symbols or abbreviations when writing Figure axis labels to avoid confusing the reader. As an example, write the quantity ÒMagnetizationÓ, or ÒMagnetization, MÓ, not just ÒMÓ. If including units in the label, present them within parentheses. Do not label axes only with units. In the example, write ÒMagnetization (A/m)Ó or ÒMagnetization {A[m(1)]}Ó, not just ÒA/mÓ. Do not label axes with a ratio of quantities and units. For example, write ÒTemperature (K)Ó, not ÒTemperature/K.Ó

\section{CONCLUSIONS}\label{sec:concl}

In this work, we demonstrated the efficacy of physics-informed neural networks in solving general sum games between swarms. Unlike conventional solvers that suffer from the curse of dimensionality as the state dimension increases, PINNs can be easily scaled with the state dimension. However, an open question still exists regarding the verification of the policies as a result of the black-box nature of the neural network, especially for the higher dimensional cases where we lack an analytical solution. In this work, we used simple case studies to demonstrate the concept of solving HJI PDEs in a swarm setting. Future works will explore applications to real-life inspired problems such as network communications. Another possible direction is the addition of an uncertainty term in the system that introduces a new layer of complexity. In the examples discussed in the paper, an implied assumption is made about the information availability. Strategies can get complicated if we introduce incomplete information in the system, which is often the case in reality. A natural extension of the present work is solving incomplete information games on swarms.

%%%%%%%%%%%%%%%%%%%%%%%%%%%%%%%%%%%%%%%%%%%%%%%%%%%%%%%%%%%%%%%%%%%%%%%%%%%%%%%%

%%%%%%%%%%%%%%%%%%%%%%%%%%%%%%%%%%%%%%%%%%%%%%%%%%%%%%%%%%%%%%%%%%%%%%%%%%%%%%%%



% \begin{thebibliography}

% \end{thebibliography}
\bibliographystyle{ieeetr}
\bibliography{ref}
\end{document}